\documentclass[preprintnumbers, floatfix,letterpaper, twocolumn,aps,prd,epsfig, nofootinbib]{revtex4-1}
\usepackage{bm,graphicx,dcolumn,epstopdf,epsf, latexsym,mathbbol, amssymb,amsmath,color,slashed, mathrsfs,mathcomp,simplewick}
\usepackage[center]{subfigure}
\usepackage{multirow}
\usepackage{makecell}
\usepackage[colorlinks,linkcolor=blue,citecolor=blue,urlcolor=blue]{hyperref}
\usepackage[center]{subfigure}
\usepackage{mathrsfs}
\usepackage[bottom]{footmisc}

\begin{document}
\allowdisplaybreaks
 \newcommand{\bq}{\begin{equation}}
 \newcommand{\eq}{\end{equation}}
 \newcommand{\bqn}{\begin{eqnarray}}
 \newcommand{\eqn}{\end{eqnarray}}
 \newcommand{\ban}{\begin{align}}
 \newcommand{\ean}{\end{align}}
  \newcommand{\nb}{\nonumber}
 \newcommand{\lb}{\label}
 \newcommand{\f}{\frac}
 \newcommand{\p}{\partial}
\newcommand{\PRL}{Phys. Rev. Lett.}
\newcommand{\PLB}{Phys. Lett. B}
\newcommand{\PRD}{Phys. Rev. D}
\newcommand{\CQG}{Class. Quantum Grav.}
\newcommand{\JCAP}{J. Cosmol. Astropart. Phys.}
\newcommand{\JHEP}{J. High. Energy. Phys.}
\newcommand{\NPB}{Nucl. Phys. B}
\newcommand{\Doi}{https://doi.org}
\newcommand{\arXiv}{https://arxiv.org/abs}
\title{ Non-adiabatic Evolution of Primordial Perturbations  and non-Gaussinity  in Hybrid Approach of Loop Quantum Cosmology}

\author{Qiang Wu${}^{a}$, Tao Zhu${}^{a}$, and  Anzhong Wang${}^{a, b}$}
\affiliation{${}^{a}$ Institute for Theoretical Physics $\&$ Cosmology, Zhejiang University of Technology, Hangzhou, 310032, China\\
${}^{b}$ GCAP-CASPER, Physics Department, Baylor University, Waco, TX 76798-7316, USA}

\date{\today}

\begin{abstract}

While loop quantum cosmology (LQC) predicts a robust quantum bounce of  the background evolution of a Friedmann-Robertson-Walker (FRW) spacetime  prior to the standard slow-roll inflation, whereby the big bang singularity is resolved,  there are several different quantization procedures to cosmological perturbations, for instance, {\em the deformed algebra, dressed metric, and hybrid quantizations}. This paper devotes to study the quantum bounce effects of primordial perturbations in the hybrid approach. The main discrepancy of this approach is the effective positive mass at the quantum bounce for the evolution of the background that is dominated by the kinetic energy of the inflaton field at the bounce, while this mass is always nonpositive in the dressed metric approach. It is this positivity of the effective mass that violates the adiabatic evolution of primordial perturbations at the initial moments of the quantum bounce. With the assumption that the evolution of the background is dominated by the kinetic energy of the inflaton at the bounce, we find that the effective potentials for both  scalar and tensor perturbations can be well approximately described by a P\"{o}schl-Teller (PT) potential, which allows us to find analytical solutions of perturbations, and from these analytical expressions  we are able to study the non-adiabatic evolution of primordial perturbations in details. In particular, we  derive their quantum bounce effects  and investigate their observational constraints. In addition, the impacts of quantum bounce effects on the non-Gaussinity and their implication on the explanations of observed power asymmetry in CMB have also been explored.


\end{abstract}


\maketitle

\section{Introduction}
\renewcommand{\theequation}{1.\arabic{equation}} \setcounter{equation}{0}

Quantization of gravity is one of the most outstanding problems in modern physics \cite{kiefer_quantum_2012}. While various approaches to quantum gravity have been pursued, including string/M-Theory \cite{becker_string_2007}, loop quantum gravity \cite{rovelli_covariant_2015}, and more recently the Horava-Lifshitz (HL) theory \cite{horava_quantum_2009} (for a recent review of the HL theory, see, for example, \cite{Wang17}), it is  fair to say that our understanding of them is still highly limited, and 
no observational evidences show  which one is correct. One of the main reasons is that the effects of quantum gravity in general appear only at the Planck scale and the corresponding quantum gravitational corrections are too small to be detectable by
 man-made terrestrial experiments in the near future.

On the other hand, the inflationary theory has become an important ingredient of modern cosmology, elegantly solving several  problems of the standard big bang cosmology \cite{guth_inflationary_1981, sato_first-order_1981, starobinsky_new_1980, baumann_tasi_2009, martin_encyclopaedia_2014}, and predicting the primordial power spectrum whose evolution determines the temperature fluctuations in the cosmic microwave background and serves as primordial seeds responding the formation of the large scale structure of our universe \cite{komatsu_seven-year_2011, planck_collaboration_planck_2014-1, planck_collaboration_planck_2015-4, planck_collaboration_planck_2018}. Considering the realization that the energy scale when inflation starts is not too far  from the Planck energy, the continuing advance in high precision cosmological observations may provide  opportunities to observe or test new fundamental physics near the Planck scale with cosmological data. Such considerations have attracted a great deal of attention in terms of understanding the quantum gravitational effects of the early Universe on the inflation and cosmological perturbations in the framework of quantum gravity, including string/M-Theory, Horava-Lifshitz theory,  loop quantum gravity/cosmology, etc. For examples, see \cite{martin_trans-planckian_2001,brandenberger_trans-planckian_2013, zhu_constructing_2014, zhu_effects_2013, zhu_detecting_2015, martineau_first_2018, ashoorioon1, ashoorioon2} and references therein.

One of the promising candidates for quantum gravity is  loop quantum gravity (LQG) \cite{rovelli_covariant_2015}. On the basis of this formalism, loop quantum cosmology (LQC) was proposed, which offers a natural framework to extend the standard slow-roll inflationary cosmology to the Planck era \cite{ashtekar_loop_2010, ashtekar_probability_2011, singh_nonsingular_2006, zhang_inflationary_2007, chen_loop_2015}. Due to the quantum gravitational effects deep inside the Planck scale, the big bang singularity arising in the inflationary scenario is replaced by a quantum bounce \cite{ashtekar_loop_2011, bojowald_quantum_2015, ashtekar_quantum_2006, ashtekar_quantum_2006-2,  ashtekar_quantum_2006-1, ashtekar_robustness_2008, yang_alternative_2009} (see also \cite{artymowski_comparison_2013, zhang_extension_2011, zhang_loop_2013, alesci_quantum_2016} for the resolution in loop quantum cosmology for modified gravity and quantum reduced loop gravity). This remarkable feature has motivated a lot of interest to consider the underlying quantum geometry effects in the background evolution of the standard inflationary scenario \cite{zhu_pre-inflationary_2017, zhu_universal_2017, li_qualitative_2018, shahalam_preinflationary_2018, shahalam_preinflationary_2018-2, shahalam_preinflationary_2017, li_towards_2018, MSWW18,agullo_pre-inflationary_2013, bonga_inflation_2016, bonga_phenomenological_2016, zhu_primoridal_2018, agullo_loop_2016,singh_2016, jin_pre-inflationary_2018}.

An important question now is whether the quantum bounce and its subsequent pre-inflationary dynamics can leave any observational signatures for the current and/or forthcoming experiments, so LQC effects can be placed directly under experimental tests. An essential step to address this issue is to implement the cosmological perturbations in the framework of LQC and calculate the corresponding inflationary observables by evolving both the scalar and tensor cosmological perturbations starting from the quantum bounce until the end of the slow-roll inflation. However, due to different quantization schemes in LQC, there are several distinct approaches to the cosmological perturbations, including deformed algebra \cite{BHKS09, MCBG12, CMBG12, CBGV12, CLB14, BBCGK15, Grain16}, group field theory \cite{gielen_cosmological_2017}, dressed metric \cite{agullo_quantum_2012, agullo_extension_2013}, and hybrid quantization approaches 
\cite{FMO12,fernandez-mendez_hybrid_2013,FMO14,gomar_cosmological_2014,gomar_gaugeinvariant_2015,CMM16,BO16,navascures_hybrid_2016}.  The evolutions of primordial perturbations during pre-inflationary phase with different quantization approaches in LQC and their footprints on primordial power spectra and non-Gaussianities have been extensively studied recently  \cite{agullo_pre-inflationary_2013, zhu_universal_2017, zhu_pre-inflationary_2017, zhu_primoridal_2018, agullo_loop_2016, singh_2016,  bonga_inflation_2016, bonga_phenomenological_2016, castellogomar_hybrid_2017, ashtekar_quantum_2017, agullo_detailed_2015, mielczarek_observational_2010, Agullo15, navascues_time-depent_2018, Grain16, bolliet_comparison_2015, schander_primordial_2016, bolliet_observational_2016}. The main characteristic of the associated effects is the evolution of perturbations during the preinflationary phase produces particles, and as a consequence the perturbations are no longer in the adiabatic BD state at the onset of the slow-roll inflation, but instead excited states. These excited states in turn produce scale-dependent features in the primordial perturbation spectrum at observable scales, and thus can be constrained by current and forthcoming observational data.

According to quantum field theory in the curved spacetime and the theory of WKB approximation, particle production can arise from non-adiabatic evolution of the associated field modes (see \cite{winitzki_cosmological_2005,agullo_non-gaussianities_2011,ashoorioon2} and references therein), which originals from the violation of the adiabatic condition of the WKB approximations. Indeed, this is exactly the case occurring for both  the cosmological scalar and tensor perturbations modes during the quantum bounce. In the deformed algebra approach, such non-adiabatic evolution and the corresponding particle productions are mainly generated when the perturbation modes evolve from the Euclidean phase to the Lorentzian phase of the quantum bounce. It is important to note that such process occurs for most modes and thus leads to significant enhancements on both scalar and tensor spectra \cite{Grain16, bolliet_comparison_2015, schander_primordial_2016}. With some reasonable assumptions and choices of initial conditions, it has been already shown that the resulting perturbation spectra are in conflict with current observations \cite{bolliet_observational_2016}. For the dressed metric and hybrid quantization approaches, the non-adiabatic evolutions are generated by an effective time-dependent mass associated with the perturbations.  The main discrepancy of the hybrid approach, as shown in \cite{navascues_time-depent_2018},  is that the effective mass is positive at the quantum bounce for the evolution of the background that is dominated by the kinetic energy of the inflaton field at the bounce, while this mass is always nonpositive in the dressed metric approach. It has been shown in the dressed metric approach that, the nonpositive mass can lead to prominent effects on primordial perturbation spectra at large scales well within current observational constraints \cite{zhu_universal_2017, zhu_pre-inflationary_2017}. In addition, these effects can also lead to an enhancement on non-Gaussianity at superhorizon scales and then provide an explanation of power asymmetry in the observational data of CMB \cite{zhu_primoridal_2018}.

Therefore, it is natural to ask if the effective time-dependent positive mass associated with perturbation modes in the hybrid approach could lead to significant footprints on primordial perturbation spectra and observational implications in explaining observational data. Earlier works on this subject have been explored by using numerical calculations \cite{castellogomar_hybrid_2017}. The purpose of this paper is to provide a detailed and analytical study of quantum gravitational effects on primordial perturbation spectra in the hybrid quantization approach and their corresponding observational implications.  More specifically, we follow the same strategies adopted in \cite{zhu_primoridal_2018, zhu_universal_2017, zhu_pre-inflationary_2017}, and show that the effective positive mass during bouncing phase can be approximated by a positive  P\"{o}schl-Teller (PT) potential. We would like to mention that this represents a distinct effect  in comparison with the negative one in the dressed metric approach. From the positive PT potential, similar to that in \cite{zhu_universal_2017, zhu_pre-inflationary_2017}, we find analytically the solutions of the perturbation modes and then calculate the corresponding quantum  effects on the primordial perturbation spectra. By using the recent released Planck 2015 data, we then obtain the observational constraint on these effects. In addition, the impacts of quantum bounce effects on the non-Gaussinity and their implication on the explanations of observed power asymmetry in CMB have also been explored.

We organize the rest of the paper as follows. In Sec. II, we provide a brief introduction to  the evolutions of background and cosmological perturbations in the hybrid quantization approach during the bouncing phase. In Sec. III, by following the same strategies used in the dressed metric approach in \cite{zhu_universal_2017, zhu_pre-inflationary_2017}, we obtain the analytical solution of perturbation mode functions and calculate explicitly the analytical expression of the corresponding power spectra. Then in Sec. IV, we use the CosmoMC code to study the observational constraints by using Planck 2015 data and the impacts of quantum bounce effects on the non-Gaussinity and their implication on the explanations of observed power asymmetry in CMB are presented in Sec. V. Our main conclusions and discussions are presented in Sec. VI.

\section{Evolution equations of both background and  perturbations during preinflationary phase}
\renewcommand{\theequation}{2.\arabic{equation}} \setcounter{equation}{0}

A robust prediction of LQC is the occurrence of a nonsingular bouncing phase, which removed the initial singularity in the early stage of the classical Universe. In this section, we present a brief introduction to  the evolution of the background and equations of motion of cosmological perturbations in the hybrid quantization approach in the preinflationary phase in LQC.

\subsection{Evolution of background  in the pre-inflationary phase}

We consider the evolution of the background for a flat Friedmann-Robertson-Walker universe with a single scalar field $\phi$. In the framework of LQC, considering the pre-inflationary period, the effective dynamics of a flat FLRW background are described by the modified Friedmann equation,
\bqn\lb{fri}
H^2=\frac{8\pi}{3 m_{\rm Pl}^2} \rho \left(1-\frac{\rho}{\rho_{\rm c}}\right),
\eqn
and the Klein-Gordon equation of the inflaton field,
\bqn
\ddot {\phi}+3H\dot{\phi}+V_{, \phi}=0,
\eqn
where $H=\dot a/a$ is the Hubble parameter with $a(t)$ being the scale factor of the FRW universe and the dot represents the derivative with respect to the cosmic time $t$, $G$ is the gravitational constant  related to the Planck mass $m_{\rm Pl}$ and the reduced Planck mass $M_{\rm Pl}$ as $8\pi G = M_{\rm Pl}^{-2} = 8 \pi m_{\rm Pl}^{-2}$, $\rho_{\rm c}$ is the critical energy density which represents the maximum value of the energy density in LQC and is about $\rho_{\rm c} \simeq 0.41 m_{\rm Pl}^4$ as suggested in black hole entropy calculations, and $V_{, \phi} = dV/d\phi$ with $V(\phi)$ being the potential of the scalar field $\phi$. The energy density $\rho$ and pressure $p$ of the scalar field are given by
\bqn
\rho&=&\frac{1}{2}\dot {\phi}^2+V(\phi), \\
p&=&\frac{1}{2}\dot {\phi}^2-V(\phi),
\eqn
with which we can define the equation of state of the scalar field as
\bqn
w_\phi \equiv \frac{p}{\rho} = \frac{\dot \phi^2-2V}{\dot \phi^2+2 V}.
\eqn
Besides the modified Friedmann and Klein-Gordon equations, it is also convenient to write the acceleration equation in the form, 
\bqn
\frac{a''}{a} = \frac{4 \pi}{3 m_{\rm Pl}^2} a^2 \rho \left(1+2 \frac{\rho}{\rho_{\rm c}}\right) - \frac{4 \pi}{m_{\rm Pl}^2} \left(1- 2 \frac{\rho}{\rho_{\rm c}}\right), ~~~~~
\eqn
where a prime denotes the derivative with respect to conformal time $\eta$  defined as $d \eta = dt /a$. It is easy to see that when $\rho$ approaches the classical  limit (i.e. $\rho \ll \rho_{\rm c}$), the above equation reduces to the standard form
\bqn\lb{acceleration_GR}
\frac{a''}{a}\simeq \frac{4 \pi}{3 m_{\rm Pl}^2} a^2 (\rho - 3 p).
\eqn

One of the remarkable consequences of Equation (\ref{fri}) is that it shows there is a quantum bounce occurring at $\rho=\rho_{\rm c}$, where the energy density reaches its maximum value and the Hubble parameter becomes zero. The background evolution with a bouncing phase has been extensively studied, and one of the main results is that, right following the quantum bounce, a desired slow-roll inflation phase is almost inevitable \cite{ashtekar_loop_2011,  singh_nonsingular_2006, zhang_inflationary_2007, chen_loop_2015, zhu_pre-inflationary_2017}. At the quantum bounce, since the energy density $\rho$ reaches $\rho_{\rm c}$, we have
\bqn
\frac{1}{2}\dot \phi_{\rm B}^2+V(\phi_{\rm B})=\rho_{\rm c},
\eqn
which imposes a strong constraint on initial conditions $(\phi_{\rm B}, \; \dot \phi_{\rm B})$ at the bounce. Among the whole $(\phi_{\rm B}, \; \dot \phi_{\rm B})$ space which satisfies the above constraints, we focus on those in which the kinetic energy dominates at the beginning (the bounce). The reason for this choice is two fold. First, for kinetic energy dominated initial conditions, the background evolution during the bouncing phase is universal and can be solved analytically \cite{zhu_universal_2017, zhu_pre-inflationary_2017}. Second, a potential dominated bounce either is not able to produce the desired slow-roll inflation because it lacks the initial kinetic energy (see examples in refs. \cite{zhu_pre-inflationary_2017, bonga_phenomenological_2016} illustrated numerically with Starobinsky potential),  or leads to a large number of e-folds of the slow- roll inflation \cite{zhu_pre-inflationary_2017, bonga_inflation_2016, bonga_phenomenological_2016, linsefors_duration_2013}. In the later case, a large number of e-folds will wash out all the observational information about the pre-inflationary dynamics and the resulting perturbations are the same as those given in GR \cite{zhu_pre-inflationary_2017, bonga_inflation_2016, bonga_phenomenological_2016}.

\begin{figure}
{\includegraphics[width=7.1cm]{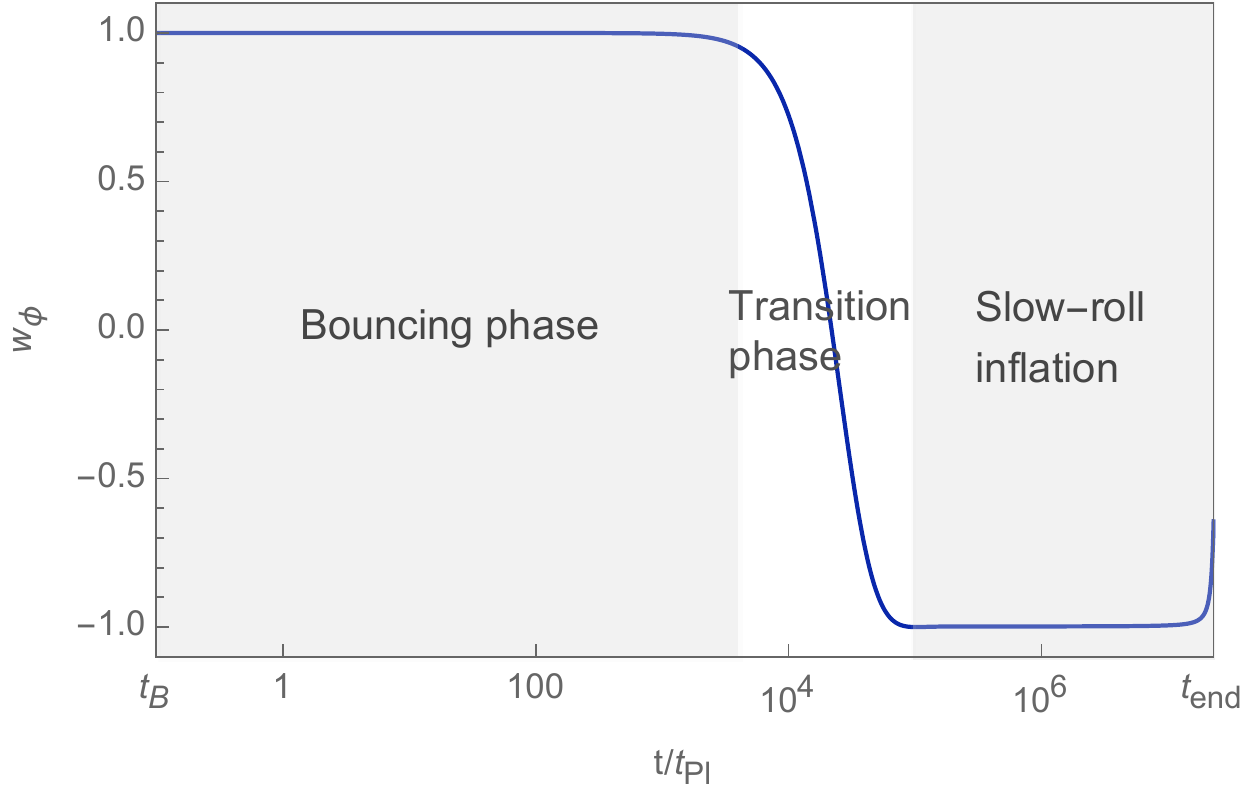}}
\caption{ Three different stages of the evolution of the Universe from the quantum bounce ($\rho=\rho_{\rm c}$) until the end of the slow-roll inflation: {\em the bouncing, transition, and slow-roll inflation phases}.}
 \label{wphi}
\end{figure}

For kinetic energy dominated initial states, the potential term in both  Friedmann and Klein-Gordon equations can be simply ignored, and then we find \cite{zhu_universal_2017}
\bqn
a(t) &=&a_{\rm B} \left(1+\gamma_{\rm B} \frac{t^2}{t_{\rm Pl}^2}\right)^{1/6},\\
\phi(t) &=& \phi_{\rm B} \pm \frac{m_{\rm Pl}}{2 \sqrt{3 \pi}}{\rm arcsinh}{\left(\sqrt{\gamma_{\rm B}} \frac{t}{t_{\rm Pl}}\right)},\\
\dot \phi(t) &=& \pm \sqrt{\frac{2 \rho_{\rm c}}{1+\gamma_{\rm B} t^2/t_{\rm Pl}^2}},
\eqn
where
\bqn
\gamma_{\rm B} \equiv \frac{24 \pi \rho_{\rm c}}{m_{\rm Pl}^4},
\eqn
is a dimensionless constant, and $t_{\rm Pl} = 1/m_{\rm Pl}$ is the Planck time. We  note that the above analytical solution is only valid when the kinetic energy  dominates. In general, the evolution of the Universe can be divided universally into three stages prior to the reheating \cite{zhu_universal_2017, zhu_pre-inflationary_2017}: {\em the bouncing, transition and slow-roll inflation}, as shown schematically in Fig. \ref{wphi} for the evolution of the equation of state $w_\phi$. With the analytical solution of $a(t)$ given above, we can obtain the relation between the conformal time $\eta$ and the cosmic time $t$ during the bouncing phase, which is
\bqn
\eta(t) - \eta_{\rm B} = t \;_2 F_1 \left(\frac{1}{6}, \frac{1}{2}, \frac{3}{2}; - \gamma_{\rm B} \frac{t^2}{t^2_{\rm Pl}}\right),
\eqn
where $\;_2 F_1 \left(\frac{1}{6}, \frac{1}{2}, \frac{3}{2}; - \gamma_{\rm B} \frac{t^2}{t^2_{\rm Pl}}\right)$ is the hypergeometric function and $\eta_{\rm B}$ denotes the conformal time at the quantum bounce.

\subsection{Equations of motion for cosmological perturbations}

There are several different approaches to implement the cosmological perturbations in the formalism of LQC, including deformed algebra, dressed metric, and hybrid quantization approaches. In this subsection, we present a brief summary about the effective equation of motion of both scalar and tensor cosmological perturbations in the hybrid approach and skip all the technical derivations of these equations. For details about the hybrid quantization of cosmological perturbations,  we refer the reader to Refs. \cite{navascures_hybrid_2016, gomar_cosmological_2014, gomar_gaugeinvariant_2015,  fernandez-mendez_hybrid_2013, navascues_time-depent_2018} and references therein.

In the hybrid quantization approach to cosmological perturbations, the effective equation of motion for the cosmological scalar and tensor perturbations are described, respectively, by \cite{navascues_time-depent_2018, navascures_hybrid_2016}
\bqn
\frac{d^2 \mu_k^{(s)}(\eta)}{d\eta^2}+\left[k^2-\frac{4\pi }{3 m_{\rm Pl}^2} a^2 (\rho-3 p) + \mathcal{U}(\eta)\right]\mu_k^{(s)}(\eta)=0,\nb\\
\lb{Muk}
\eqn
and
\bqn
\frac{d^2 \mu_k^{(t)}(\eta)}{d\eta^2}+\left[k^2-\frac{4\pi }{3 m_{\rm Pl}^2} a^2 (\rho-3 p) \right]\mu_k^{(t)}(\eta)=0,\nb\\
\lb{Mukt}
\eqn
where
\bqn
\mathcal{U}(\eta)=a^2\left[V_{,\phi\phi}+48\pi GV+6\frac{a'\phi'}{a^3\rho}V_{,\phi}-\frac{48\pi G}{\rho}V^2\right], \nb\\
\eqn
$\mu_k^{(s, t)}$ denotes the Mukhanov-Sasaki variable with $\mu^{(s)}_k(\eta)=z_s \mathcal{R}_k$ and $\mu^{(t)}_k(\eta) = a h_k/2$ where $\mathcal{R}_k$ denotes the comoving curvature perturbations, $h_k$ the tensor perturbations, and $z_s =a\dot \phi/H$.

Since we only focus on the background evolution with kinetic energy dominated initial conditions, during the bouncing phase, $\mathcal{U}(\eta)$ in (\ref{Muk}) is negligible \cite{zhu_universal_2017, zhu_pre-inflationary_2017}. As a result, the equations of motion for the cosmological scalar and tensor perturbations approximately take the same form during the bouncing phase. During the transition phase, the energy density $\rho$ drops down to about $10^{-12} \rho_{\rm c}$, and then we can ignore all the LQC corrections in the equations of motion. Hence, from Eq.~(\ref{acceleration_GR}) we find, 
\bqn
\frac{4\pi}{3 m_{\rm Pl}^2} a^2 (\rho-3 p)- \mathcal{U}(\eta) \to \frac{a''}{a}-\mathcal{U}(\eta) \simeq \frac{z_s''}{z_s},
\eqn
and
\bqn
\frac{4\pi}{3 m_{\rm Pl}^2} a^2 (\rho-3 p) \to \frac{a''}{a},
\eqn
so thereafter both the scalar and tensor perturbations enter into the classical regime and their perturbation equations reduce precisely to those obtained in GR.

\subsection{Time-dependent and positive effective mass}

As shown in \cite{navascues_time-depent_2018}, one of the main characteristics of cosmological perturbations in the hybrid quantization approach is that the  time-dependent effective mass of perturbation modes is positive near by   the quantum bounce. To see this clearly, let us define the effective mass function for both scalar and tensor perturbations as
\bqn
m_{\rm eff}^2(\eta) = - \frac{4 \pi }{3 m_{\rm Pl}^2}a^2 (\rho-3p),
\eqn
where for scalar perturbations we ignore  the contributions from the term $\mathcal{U}(\eta)$ since we only focus on kinetic energy dominated initial conditions, for which  $\mathcal{U}(\eta)$ is very small during the whole bouncing phase \cite{zhu_universal_2017, zhu_pre-inflationary_2017}. Therefore, the effective mass function $m_{\rm eff}^2$ takes the same form for both scalar and tensor perturbations during the bouncing phase.

\begin{figure*}
{\includegraphics[width=7.0cm]{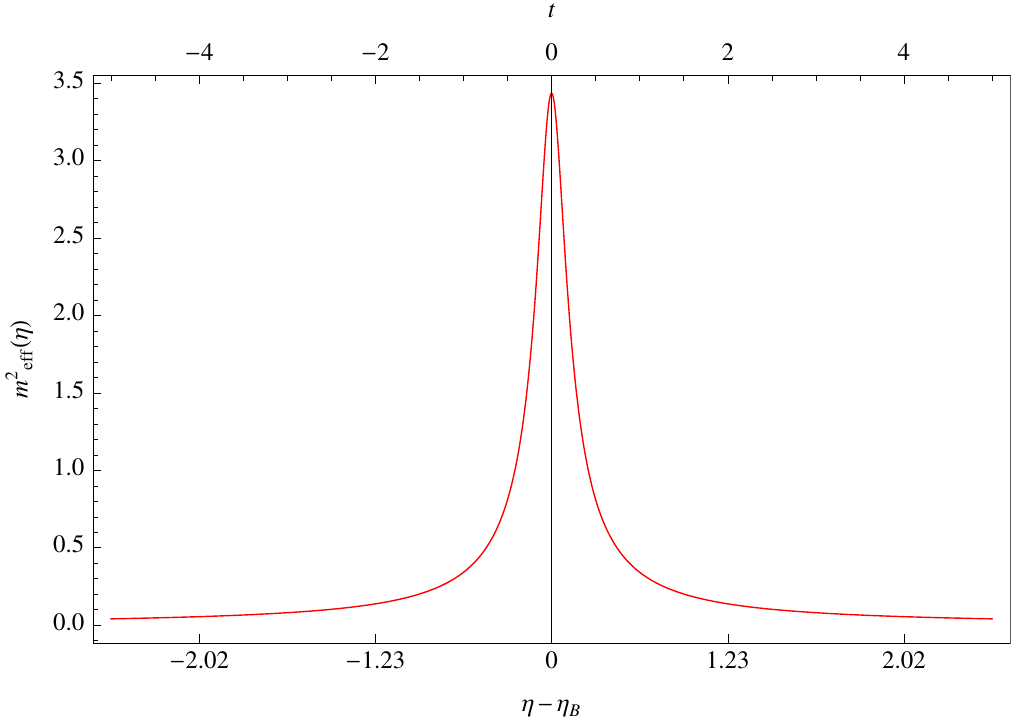}}
{\includegraphics[width=7.0cm]{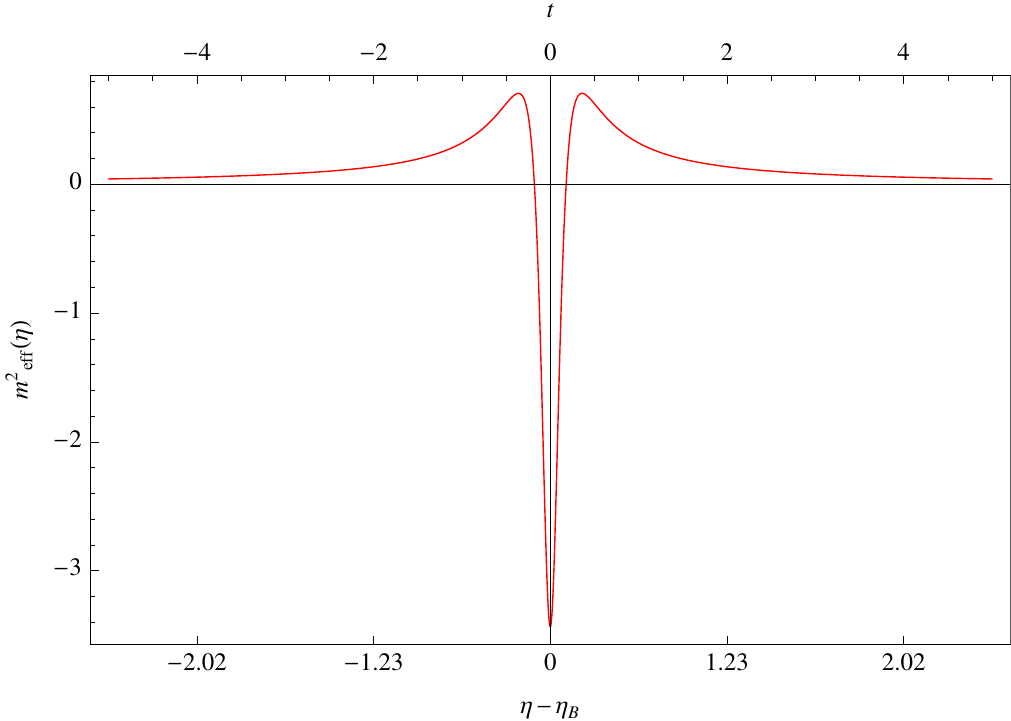}}
\caption{Comparison of the effective time-dependent mass function $m_{\rm eff}^2(\eta)$ of field equations of cosmological perturbation modes in the hybrid quantization and dressed metric approaches during bouncing phase. The left panel presents the effective mass function in the hybrid quantization approach while the right panel presents that in the dressed metric approach.}
 \label{effective_mass}
\end{figure*}

Using the analytical solutions for $a(t)$ and $\phi(t)$ obtained in the above subsection, the effective mass function can be casted into the form
\bqn\lb{mass_hybrid}
m_{\rm eff}^2(\eta) = \frac{a_{\rm B}^2 \gamma_{\rm B} m_{\rm Pl}^2}{9} \left(1+\gamma_{\rm B} \frac{t^2}{t^2_{\rm Pl}}\right)^{-2/3},
\eqn
which shows explicitly that the effective mass for both scalar and tensor perturbations are always positive during the whole bouncing phase. This property represents a distinguishing characteristic of the hybrid quantization approach, in comparison with other quantization approaches to the cosmological perturbations in LQC. For example, in the dressed metric approach, as shown in \cite{zhu_universal_2017, zhu_pre-inflationary_2017}, the effective mass function  can be expressed as
\bqn\lb{mass_dressed}
m_{\rm eff}^2(\eta) = - \frac{a''}{a} = - a_{\rm B}^2 \frac{\gamma_{\rm B} m_{\rm Pl}^2 (1- \gamma_{\rm B} t^2/t^2_{\rm Pl})}{9 (1+\gamma_{\rm B} t^2/t^2_{\rm Pl})^{5/3}}.
\eqn
Obviously, this effective mass exhibits different behavior around the quantum bounce where it becomes negative for $-1/\sqrt{\gamma_{\rm B}}<t < 1/\sqrt{\gamma_{\rm B}}$. Fig.~\ref{effective_mass} displays the  time-dependent mass  $m_{\rm eff}^2$ in both hybrid quantization and dressed metric approaches, which shows clearly the difference of the mass functions in both approaches. It has been explained in \cite{navascues_time-depent_2018} that the discrepancy in hybrid quantization and dressed metric approaches appears only in the  time-dependent effective mass term of the corresponding field equations during the bouncing phase, and the origin of this difference arises from the distinct quantization procedures.

\subsection{Violations of adiabaticity of cosmological perturbations during   bouncing phase}

To study the evolution of perturbation modes with the effective time-dependent mass, a convenient way is to use the WKB analysis. In general, the solution of the mode function $\mu_k^{(s,t)}(\eta)$ of the equation,
\bqn
\frac{d^2 \mu_k^{(s, t)}(\eta)}{d\eta^2} + \Omega^2(\eta) \mu_k^{(s,t)}(\eta)=0,
\eqn
can be approximately given in terms of the WKB solutions
\bqn\lb{eom_wkb}
\mu_k^{(s, t)}(\eta) \simeq \frac{\alpha_k}{\sqrt{2 \Omega(\eta)}}e^{- i \int \Omega(\eta) d\eta} + \frac{\beta_k}{\sqrt{2 \Omega(\eta)}}e^{ i \int \Omega(\eta) d\eta},\nb\\
\eqn
if the WKB condition, or JWKB criterion
\bqn \lb{wkb}
\left|\frac{3 \Omega' {^2}}{4\Omega^4}-\frac{\Omega''}{2\Omega^3}\right|\ll 1,
\eqn
is satisfied. Here the function $\Omega^2(\eta) \equiv k^2 + m_{\rm eff}^2(\eta)$, and $\alpha_k$ and $\beta_k$ are two Bogoliubov coefficients, which can be determined by choosing an initial state of the modes. Generally an adiabatic state is a good choice, and if the WKB condition is satisfied during the whole process, we have
\bqn
\alpha_k=1, \; \beta_k =0.
\eqn
However, in some cases, the WKB condition may be violated or not be fulfilled completely. Then, the non-adiabatic evolution of the mode $\mu_k^{(s, t)}(\eta)$ will produce excited states (i.e. particle production) during this process and eventually lead to a state with
\bqn
\alpha_k \neq 1,\; \beta_k \neq 0.
\eqn

According to (\ref{wkb}), there are several facts that can lead to the violation of the WKB condition. One case is that $\Omega^2(\eta)$ contains zeros (represented as real turning points of Eq. (\ref{eom_wkb})) or extremely closes to zero (complex conjugated turning points of Eq.~(\ref{eom_wkb})) in the intervals of interest. It is simple to check that when $\Omega^2(\eta)$ equals zero, the WKB condition in (\ref{wkb}) becomes divergent. Another possible case that could violate the WKB condition is around the extreme point of the function $\Omega^2(\eta)$. Since $\Omega'(\eta)=0$ at the extreme point, it is easy to see that the WKB condition can be violated if
\bqn
\left|\frac{\Omega_m''}{2 \Omega_m}\right|>1,
\eqn
where $\Omega''_m$ and $\Omega_m$ represent the value of $\Omega''(\eta)$ and $\Omega(\eta)$ at the extreme point.

\begin{figure*}
{\includegraphics[width=8.1cm]{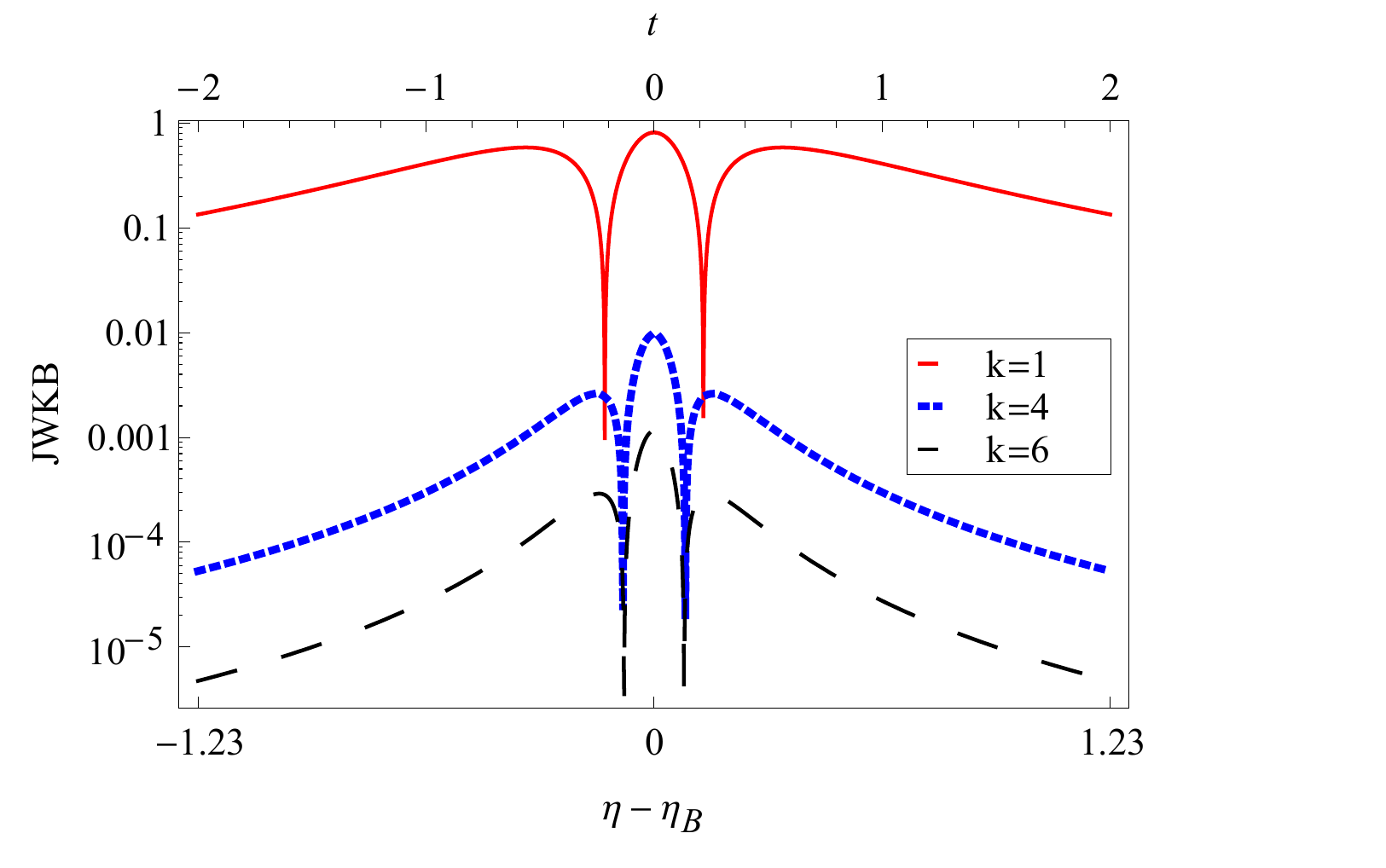}}
{\includegraphics[width=8.1cm]{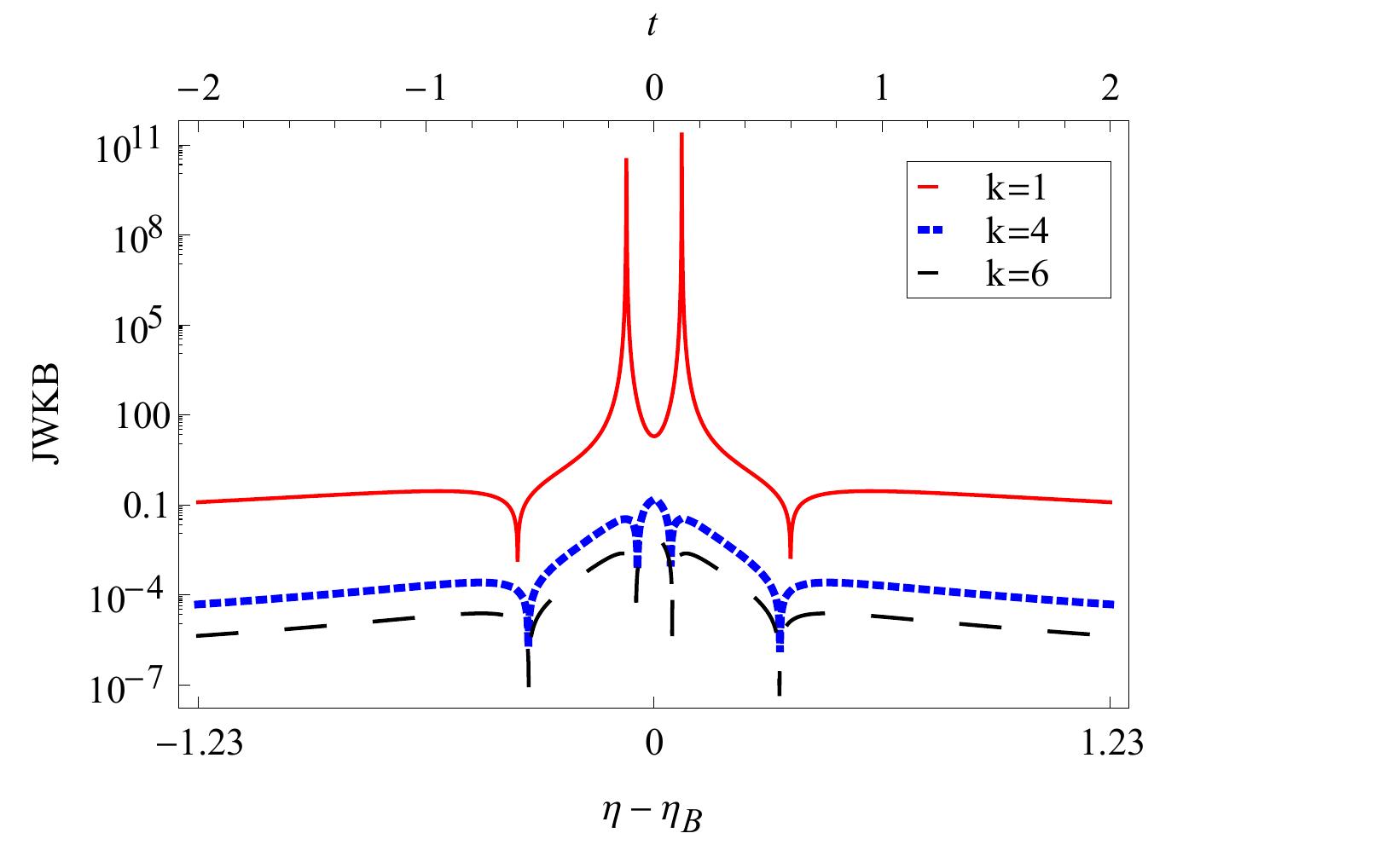}}
\caption{JWKB criterion is violated near the time of bounce at $t=0$. The left panel shows the result for the hybrid approach and the right panel shows the result for the dressed metric approach. Note that we used unit $m_{\rm Pl}=1$ and set $a_{\rm B}=1$ in these figures.}
 \label{jwkb}
\end{figure*}

Now the question is whether the  effective time-dependent and positive mass given in (\ref{mass_hybrid}) in the hybrid quantization approach could lead to the violation of the WKB approximation. Because of the positivity of this effective time-dependent mass, the function $\Omega^2(\eta) = k^2+m^2_{\rm eff}(\eta)$ is always positive and $\Omega^2(\eta)$ does not contains zeros during the whole bouncing phase. Now we can check if the extreme point of $\Omega^2(\eta)$ could lead to violation of the WKB condition. We observe that the extreme point of $\Omega^2(\eta)$ locates exactly at the quantum bounce, i.e., $t=0$ ($\eta=\eta_{\rm B}$), and we have
\bqn\lb{wkb_B}
\left|\frac{\Omega''(\eta_{\rm B})}{2 \Omega^3(\eta_{\rm B})}\right| = 3 \left(1+\frac{k^2}{k_{\rm H}^2}\right)^{-2},
\eqn
where
\bqn
k_{\rm H} \equiv \frac{\sqrt{\gamma_{\rm B}} a_{\rm B} m_{\rm Pl}}{3},
\eqn
which defines a characteristic energy scale of hybrid quantization approach in LQC. Note that we use $k_{\rm H}$ to distinguish the energy scale $k_{\rm B}$ of dressed metric approach in LQC \cite{zhu_universal_2017}. From Eq.~(\ref{wkb_B}), it is obvious that the WKB condition is well satisfied if $k \gg k_{\rm H}$, and it is violated when $k \lesssim k_{\rm H}$. This property is in agreement with the behaviors of the WKB condition presented in the left panel of Fig.~\ref{jwkb} for different values of wavenumber $k$. Therefore, for perturbation modes with large wavenumber $k$ the adiabatic evolution always hold during the whole bouncing phase, while the adiabaticity is violated at quantum bounce for modes with small wavenumber.

One may also be interested in the difference between the hybrid quantization and dressed metric approach in the above WKB analysis. In the dressed metric approach, as shown in (\ref{mass_dressed}), the effective time-dependent mass is negative at the quantum bounce. With this fact, the violations of the WKB condition in the dressed metric approach may come from two parts for certain modes. First, the function $\Omega^2(\eta)$ could contain zeros or complex conjugated zeros when $k$ is small, and in this case, the WKB condition becomes divergent or large enough. Second, similar to that in hybrid quantization approach, at the quantum bounce,
\bqn
\left|\frac{\Omega''(\eta_{\rm B})}{2 \Omega^3(\eta_{\rm B})}\right|=\frac{21}{2}\left(3+\frac{k^2}{k^2_{\rm B}}\right)^{-2},
\eqn
where $k_{\rm B} \equiv a_{\rm B} m_{\rm Pl} \sqrt{\gamma_{\rm B}/3}$ denotes a characteristic energy scale of dressed metric approach in LQC. Obviously, the WKB condition is violated for modes with $k \lesssim k_{\rm B}$ while it is fulfilled for modes with $k \gg k_{\rm B}$. In the right panel of Fig.~\ref{jwkb}, we display the WKB condition in the dressed metric approach for several different values of $k$. Whereas the WKB condition can only be violated at quantum bounce for some $k$ in the hybrid quantization approach, Fig.~\ref{jwkb} shows clearly that it now may be violated around the zeros and the extreme point (i.e. at the quantum bounce) of $\Omega^2(\eta)$.

\section{Non-adiabatic effects on primordial perturbations spectra and its observational constraints}
\renewcommand{\theequation}{3.\arabic{equation}} \setcounter{equation}{0}

\subsection{Analytical solution with PT potential}

According to the analysis presented in the above section, the adiabaticity of perturbation modes with $k \gg k_{\rm H}$ is always fulfilled during the whole bouncing phase. For these modes, if one chooses an adiabatic initial state, they remain at the adiabatic state until they exit the Hubble horizon during the slow-roll inflation. Therefore, the quantum bounce effects of the hybrid quantization approach are suppressed and the corresponding perturbation spectra for both scalar and tensor modes take precisely the standard power-law form as that given in general relativity. The more interesting modes are those that violate the WKB condition and thus experience a non-adiabatic evolution during the  bouncing phase, which in turn produces excited states  as we mentioned in the above section. Now an essential step is to estimate the non-adiabatic effects on perturbation spectra by solving the equations of motion during the bouncing phase.

\begin{figure}
{\includegraphics[width=8.1cm]{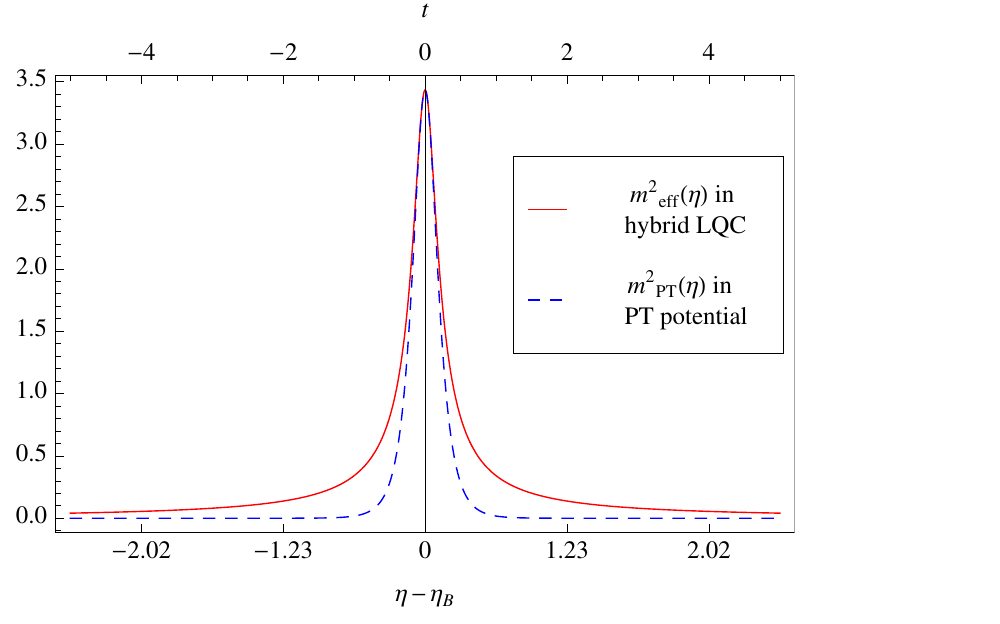}}
\caption{Comparison between time-dependent mass   $m^2_{\rm{eff}}(\eta)$ in the hybrid quantization and the PT potential $m^2_{\rm{PT}}(\eta)$. }
 \label{fig_pt}
\end{figure}

For this purpose, we follow the same strategy used in \cite{zhu_universal_2017, zhu_pre-inflationary_2017} for perturbation modes in the dressed metric approach, in which the effective time-dependent mass is approximated by an effective PT potential with a negative sign. For perturbation modes we are interested in in the hybrid quantization approach, since the effective time-dependent mass is always positive, we can use a PT potential with positive sign to describe it.  This PT potential takes the form
\bqn
m_{\rm PT}^2(\eta)= \frac{V_0}{{\rm cosh}^2\alpha (\eta-\eta_{\rm B})},
\lb{pt}
\eqn
where $V_0= a_{\rm B}^2 m_{\rm Pl}^2 \gamma_B/9$ and $\alpha^2=2 a_{\rm B}^2 m_{\rm Pl}^2 \gamma_B /3$. We note that in the dressed metric approach, the effective mass is described by a PT potential $m_{\rm PT}^2(\eta) = - V_0{\rm cosh}^{-2} [\alpha (\eta-\eta_{\rm B})]$ with $V_0 = a_{\rm B}^2 m_{\rm Pl}^2 \gamma_B/3$ and $\alpha^2 = 2 a_{\rm B}^2 m_{\rm Pl}^2 \gamma_B $.
 We plot the PT potential and effective mass term in Fig. \ref{fig_pt} where we can see that they match each other well around the quantum bounce at $t=0$.

 With the PT potential, we find the analytical solution of Eq. (\ref{Muk}), which  has the general form,
\bqn\lb{sol_PT}
\mu^{(\text{PT})}_k(\eta) &=& a_k x^{ik/(2\sqrt{6} k_{\rm H})} (1-x)^{-ik/(2 \sqrt{6} k_{\rm H})} \nb\\
&& \times \; _2F_1(a_1-a_3+1,a_2-a_3+1,2-a_3,x)\nb\\
&& +b_k [x (1-x)]^{-ik /(2 \sqrt{6} k_{\rm H})} \;_2F_1(a_1,a_2,a_3,x),\nb\\
\lb{ptsolution}
\eqn
Where
\bqn
x(\eta) &=& \frac{1}{1+e^{-2 \alpha (\eta-\eta_\text{B})}}, 
\eqn
and
\bqn
\lb{acs}
a_1 &\equiv& \frac{1}{2}\left(1+\frac{\sqrt{15}}{6}\right)  -\frac{i k}{\alpha},\nb\\
a_2 &\equiv & \frac{1}{2}\left(1-\frac{\sqrt{15}}{6}\right) -\frac{i k}{\alpha},\nb\\
a_3 &\equiv & 1-\frac{i k}{\alpha}.
\eqn
In the above solution, $a_k$ and $b_k$ are two independent integration constants which should be uniquely determined by the initial conditions.

\subsection{Primordial perturbation spectrum with quantum gravitational  effects}

We need  to fix the initial state in order to determine   the integration constants $a_k$ and $b_k$. One of the choices  to impose the initial conditions is at the quantum bounce. However, this is a subtle issue since adiabaticity of some modes is violated and it seems impossible to impose an adiabatic state at this point \cite{agullo_unitarity_2015, agullo_preferred_2015, ashtekar_initial_2017, zhu_pre-inflationary_2017}. Another choice that has been frequently used is the remote past \cite{MBS17}. For this choice, one needs to analytically extend the bouncing phase to the contracting phase right before the quantum bounce. In the contracting phase, as shown in Fig.~\ref{jwkb}, the adiabatic conditions of the perturbation modes are fulfilled and therefore we can choose an adiabatic state as the initial state of perturbation modes, which takes the form
\bqn
\mu^{initial}_{k}(\eta) \sim \frac{1}{\sqrt{2k}}e^{-ik \eta},
\eqn
with which we can uniquely determine the coefficients $a_k$ and $b_k$ in Eqs.(\ref{ptsolution}). Here we would like to mention that in both the above adiabatic vacuum state and the analytical solution with PT potential, we have assumed that the effective mass term is negligible for both scalar and tensor modes at the initial time where one specifics adiabatic state. With this assumption we have to require that the wavenumber $k^2 \gg |a''/a|$ at that time. For the modes with very small wavenumber $k^2 \lesssim |a''/a|$ initially, the initial state has to be chosen with cautions. Another strategy for imposing initial state in the contracting phase is studied in Ref.~\cite{barrau_scalar_2018}, in which the initial states are imposed at the time when the effective time-dependent mass term vanishes. For large wavenumber modes their choice recovers the usual adiabatic states in the contracting phase and for small wavenumber modes the final perturbation spectrum may be different for initial states imposed at different zeros of the effective time-dependent mass.

\begin{figure}
{\includegraphics[width=8.1cm]{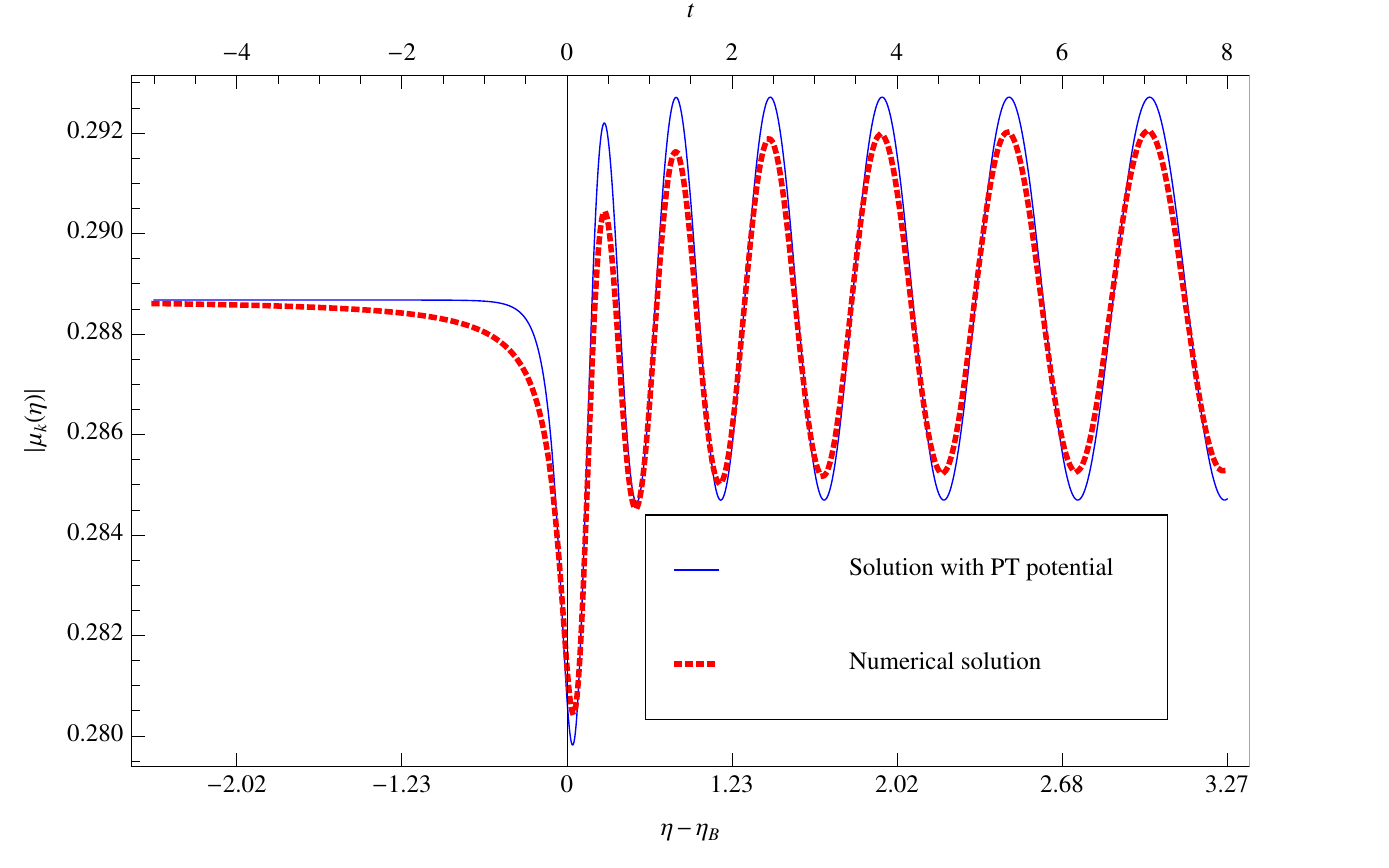}}
\caption{Comparison between the analytical solution and numerical one with $a_{\rm B}=1$, $m_{\rm Pl}=1$ and $k=6$.}
 \label{solk6}
\end{figure}

Using the asymptotic expressions of the hypergeometric functions when $\eta-\eta_B\ll0$, i.e.,
\bqn
&& x \sim e^{2\alpha (\eta-\eta_B)}\rightarrow 0,\nb \\
&& x^{ik/(2\alpha)}(1-x)^{-ik/(2\alpha)}\sim e^{\alpha (\eta-\eta_B)},
\eqn
and the fact $_2F_1(c_1,c_2,c_2,0)=1$, we find
\bqn
a_k=0,\,\,\,\  b_k=\frac{e^{ik\eta_B}}{\sqrt{2k}}.
\eqn
Then, the evolutions of both scalar and tensor perturbations are completely determined. In Fig.~\ref{solk6}, we  display the analytical solution comparing with the  numerical one with the values of $a_{\rm B}=1$,$m_{\rm Pl}=1$ and $k=6$, which shows that they match very well during the bouncing phase.

After the bouncing phase and before the slow-roll inflationary phase, the perturbations evolve into the transition phase with $k^2\gg m^2_{\rm eff}(\eta)$, during which the perturbation modes have the general solution,
\bqn
\mu_k^{\alpha}(\eta)=\frac{1}{\sqrt{2k}}(\tilde{\alpha}_ke^{-ik\eta}+\tilde{\beta}_k e^{ik\eta}).
\lb{soltransition}
\eqn
We emphasize here that this solution is also valid during the slow-roll inflation until the modes exit the Hubble horizon, thus it can be matched to the standard analytical solution during the slow-roll inflation,
\bqn
\mu^{(sr)}_k(\eta)\simeq \frac{\sqrt{-\pi\eta}}{2}[\alpha_k H^{(1)}_{\nu}(-k\eta)+\beta_k H^{(2)}_{\nu}(-k\eta)],\nb\\
\lb {solslowroll}
\eqn
where $\nu$ is assumed to be a constant during the slow-roll inflation and $H_{\nu}^{(1,2)}$ represent the Hankel functions of the first and second kind respectively.

\begin{figure}
{\includegraphics[width=8.1cm]{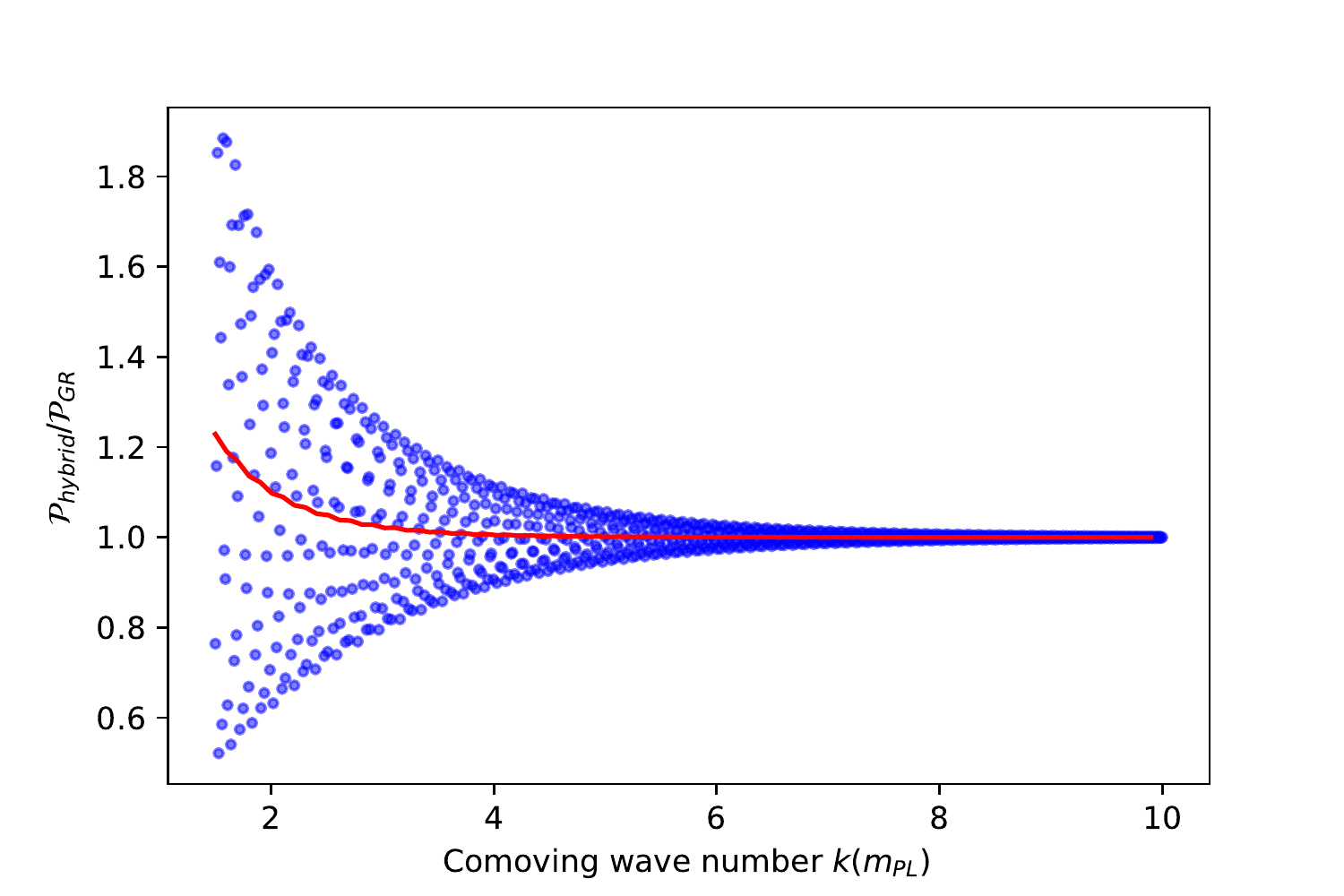}}
\caption{The primordial power spectrum obtained with analytic solutions for the perturbations. The oscillating points show the power spectrum computed for each mode and the red line shows the binned average of these points. We set $\eta_B=2000 $ in this figure.}
 \label{fig_p}
\end{figure}

Then, we can match the solutions given by  Eqs.(\ref{soltransition}), (\ref{solslowroll}) and (\ref{ptsolution}) at the transition phase for $\eta-\eta_{\rm B} \gg 0$, i.e.,
\bqn
x\sim 1-e^{-2\alpha(\eta-\eta_B)}\rightarrow1,
\eqn
and find
\bqn\lb{akbk}
\alpha_k&=&\frac{\Gamma(a_3)\Gamma(a_1+a_2-a_3)}{\Gamma(a_1)\Gamma(a_2)}e^{2ik\eta_{\rm B}}, \nb \\
\beta_k &=&\frac{\Gamma(a_3)\Gamma(a_3-a_1-a_2)}{\Gamma(a_3-a_1)\Gamma(a_3-a_2)}.
\eqn
In GR, one in general imposes the adiabatic vacuum state when the modes are inside the Hubble horizon, i.e., $\alpha_k=1,\; \beta_k=0$. However, we show clearly that if there is a bouncing phase prior to the slow-roll inflation, $\beta_k$ now does not vanish generically. This leads to modifications at the onset of the slow-roll inflation on the standard nearly scale invariant power spectrum \footnote{For the standard nearly scale invariant power spectrum $\mathcal{P}_{\mathcal{R}}^{\rm GR} =  \frac{H^2}{8 \pi^2 m_{\rm Pl}^2 \epsilon_1}$, we have assumed that all the relevant modes at observable scales exit the Hubble horizon at the time when the slow-roll approximation is fulfilled.  But this assumption can be violated at large scales if the number of e-folds between the bounce and the slow-roll inflation is not large enough, for which the relevant modes at large scales could exit the Hubble horizon even when the slow-roll approximation is not yet completely satisfied. We would like to note that even for these modes the expression (\ref{pw}) is still valid but suppressed by a large value of $\epsilon_1$ in comparing to those at smaller scales.} ,
\bqn\lb{pw}
\mathcal{P}_{\mathcal{R}}(k) = \left|\alpha_k+\beta_k\right|^2 \frac{H^2}{8 \pi^2 m_{\rm Pl}^2 \epsilon_1},
\eqn
where
\bqn\lb{akbk2}
|\alpha_k&&+\beta_k|^2=1+\left[1+\cos \left(\sqrt{\frac{5}{3}} \pi \right)\right] \text{csch}^2\left(\frac{\pi  k}{\alpha }\right) \nb\\
                    &&+\sqrt{2} \sqrt{\cosh \left(\frac{2 \pi  k}{\alpha }\right)+\cos \left(\sqrt{\frac{5}{3}} \pi \right)}\left|\cos\left(\frac{1}{2} \sqrt{\frac{5}{3}} \pi \right)\right|\nb\\
                    &&\times \text{csch}^2\left(\frac{\pi  k}{\alpha }\right)\cos\left(2k\eta_B+\varphi_k\right),
\eqn
with
\bqn\lb{pw_alpha}
\varphi_k\equiv\arctan\left\{\frac{\text{Im}[\Gamma(a_1)\Gamma(a_2)\Gamma^2(a_3-a_1-a_2)]}{\text{Re}[\Gamma(a_1)\Gamma(a_2)\Gamma^2(a_3-a_1-a_2)]}\right\}.\nb\\
\eqn
In Fig.~\ref{fig_p} we display the ratio between the power spectrum with the bouncing effects and the standard one given in GR, as a function of wavenumber $k$. We would like to mention that, the behavior of the power spectrum in Fig.~\ref{fig_p} is in agreement with that given in \cite{castellogomar_hybrid_2017}. While the results obtained in \cite{castellogomar_hybrid_2017} are purely numerical, here ours are derived directly from the analytical expression of Eq.~(\ref{pw}).

Another important property of the bouncing effects is that they result in scale-dependent oscillations in the perturbation spectrum. The amplitudes of these oscillations essentially depend on the parameter $k_{\rm H}$, thus represent a characteristic feature of LQC. In Eq.~(\ref{akbk2}), the last two terms, proportional to ${\rm csch}^2(\pi k/(\sqrt{6} k_{\rm H}))$, increase exponentially as $k$ decreases. This implies that the quantum gravitational  effects on the perturbation spectrum get enhanced for $k\lesssim k_{\rm H}$, and suppressed for $k \gg k_{\rm H}$. In the latter case, the perturbation spectrum reduces to the standard power-law spectrum in GR. Therefore, the quantum gravitational effects are important at the scales $k \lesssim k_{\rm H}$. This is in agreement with the qualitatively analysis in the above section, where the modes with $k \lesssim k_{\rm H}$ violate the WKB condition at the quantum bounce,  and thus lead to significant changes in the primordial perturbation spectrum.

In addition, it is also worthwhile to note that the properties of quantum  effects in the primordial perturbation spectra  in both hybrid and dressed metric approaches are qualitatively the same. Quantitatively, the effects in both approaches are different in two aspects. First, these two approaches have two different characteristic energy scales, i.e, $k_{\rm H}$ in the hybrid quantization approach and $k_{\rm B}$ in the dressed metric approach. Second, as we can seen from Eq.~(\ref{akbk2}), the numerical factors in front of ${\rm csch}^2(\pi k/(\sqrt{6} k_{\rm H}))$ are in general different from those obtained in \cite{zhu_universal_2017, zhu_pre-inflationary_2017} in the dressed metric approach.

\subsection{Observational constraints}

\begin{figure*}
\includegraphics[width=8.1cm,height=5.5cm]{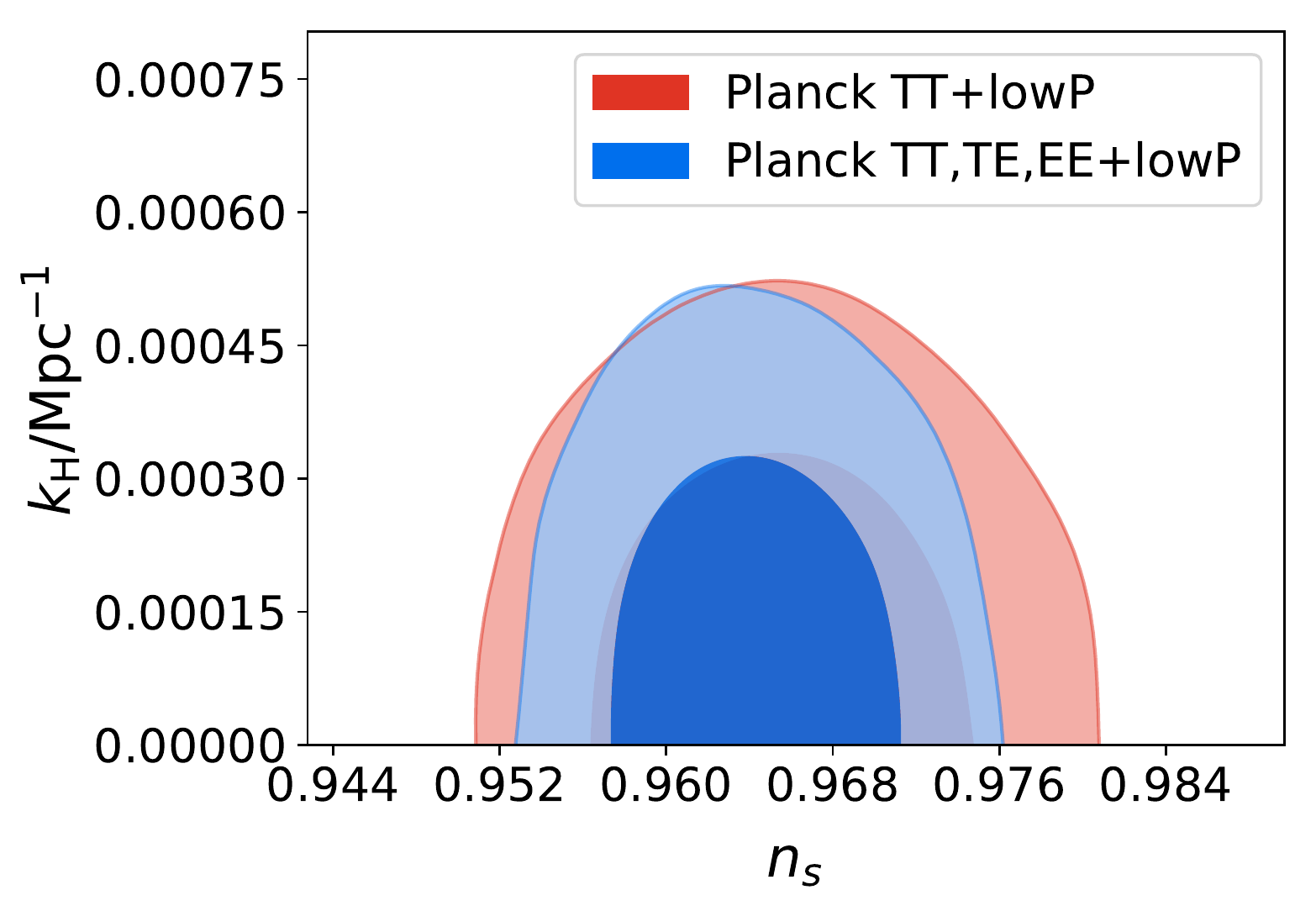}
\includegraphics[width=8.1cm,height=5.5cm]{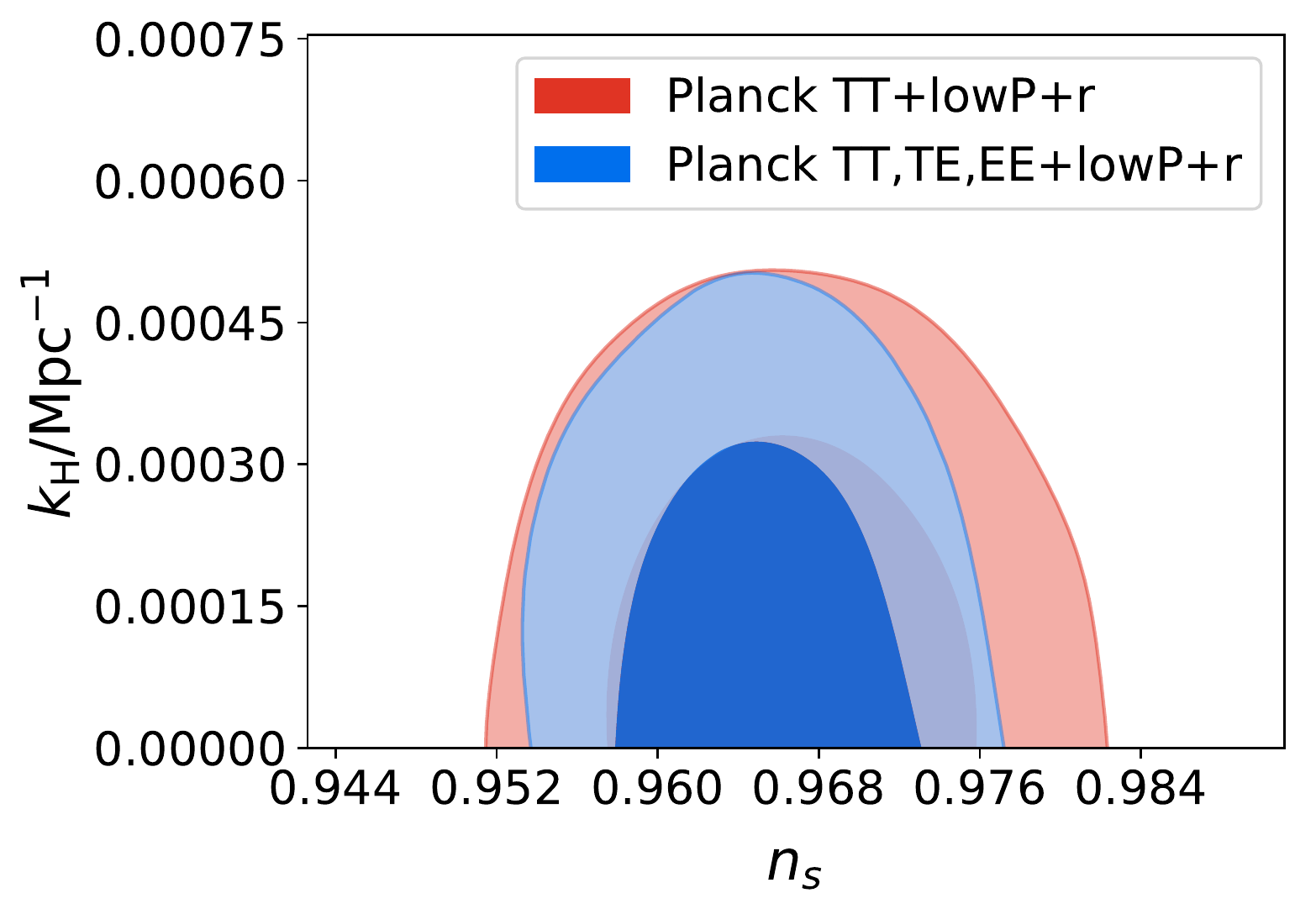}
\includegraphics[width=8.1cm,height=5.5cm]{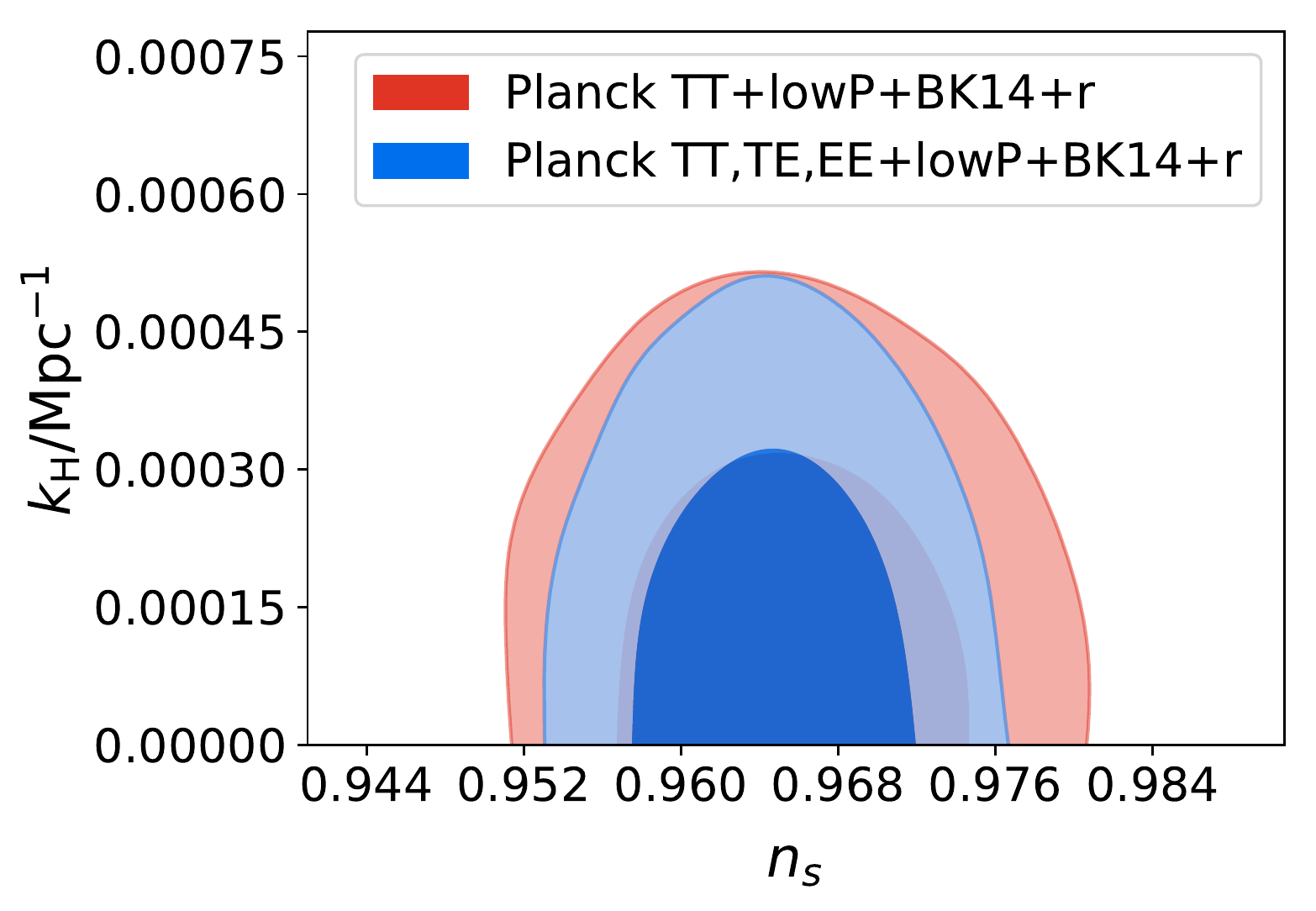}
\caption{The CMB likelihood analysis  in the ($k_H,\; n_s$)-plane with a robust fitting  $n_s \simeq 0.965$.
The observational constraints on $(n_s, k_\text{H}/\text{Mpc}^{-1})$ are obtained at 68\% and 95\% C.L. by using Planck 2015 TT+lowP, Planck 2015 TT, TE, EE+lowP, and BIECP/KECK 2014 data. The left upper panel only considers the scalar spectrum, while the right upper panel and the bottom one includes the non-zero tensor contributions.  Note that we set $a_0=1$. } \label{CMB_Likelihood}
\end{figure*}

In this subsection, we perform the CMB likelihood analysis by using the Planck 2015 data, with the MCMC code developed in \cite{lewis_cosmological_2002}. For this purpose we parametrize the  primordial perturbations spectra as
\bqn
\mathcal{P}_{\mathcal{R},h}(k)= |\alpha_k&&+\beta_k|^2\mathcal{P}^{GR}_{\mathcal{R},h}(k),
\eqn
where
\bqn
\mathcal{P}^{GR}_{\mathcal{R}}(k)&=& A_s\left(\frac{k}{k_*}\right)^{n_s-1+\cdots}\nb \\
\mathcal{P}^{GR}_{h}(k)&=& A_t\left(\frac{k}{k_*}\right)^{n_t+\cdots}.
\eqn
Here $A_s (A_t)$ is the scalar (tensor) amplitude, $n_s (n_t)$  the scalar (tensor) spectral index, and $k_*=0.05 {\rm Mpc}^{-1}$ denotes the pivot scale.

We assume the flat cold dark matter model with the effective number of neutrinos $N_{\text{eff}}=3.046$ and fix the total neutrino mass $\Sigma m_\nu=0.06 eV$.
Let us first  consider the scalar spectrum and vary the following seven parameters,
\bqn
(\Omega_{\rm b}h^2, \Omega_\text{c}h^2, \tau, \Theta_s, n_s, A_s, k_\text{H}/a_0),
\eqn
where $\Omega_{\rm b}h^2$ and $\Omega_\text{c}h^2$ are, respectively,  the baryon and cold dark matter densities, $\tau$ is the optical depth to reionization, $\Theta_s$ is the ratio (multiplied by $100$)
of the sound horizon at decoupling to the angular diameter distance to the last scattering surface. In addition, we have one more parameter $k_\text{H}/a_0$, which is related to the effects of the pre-inflationary dynamics.
For the six cosmological parameters ($\Omega_{\rm b}h^2, \Omega_\text{c}h^2, \tau, \Theta_s, n_s, A_s$), we use the same prior ranges as in \cite{planck_collaboration_planck_2013}, while for the parameter $k_{\rm H}/a_{0}$,
which is related to the bouncing effects, we set the prior range as $k_{\rm H}/a_0 \in [10^{-8}, 0.002] {\rm Mpc}^{-1}$.

In particular, we use the high-$l$ CMB temperature power spectrum (TT) and the polarization data (TT, TE, EE) respectively with low-$l$ polarization data (lowP) from Planck 2015, and BK 14 from BICEP/KECK 2014 data. In Table.~\ref{bestfit}, we list the best fit values of the six cosmological parameters and constraints on $k_{\rm H}/a_0$ and $r$ at $95\%$ C.L. for different cosmological models from different data combinations.
 Marginalizing other parameters, we find that $k_\text{H}/a_0$ is constrained by the Planck TT+lowP (Planck TT,TE,EE+lowP) to 
\bqn
\frac{k_\text{H}}{a_0} < 4.17 \times 10^{-4}\text{Mpc}^{-1} (4.15 \times 10^{-4}), \;\; {\rm at \; 95\% \; C.L}.\nb\\
\eqn
When we add one more parameter, the tensor-to-scalar ratio $r=A_{(t)}/A_{(s)}$, to include the tensor spectrum, the Planck TT+lowP (Planck TT,TE,EE+lowP) data yields 
\bqn
\frac{k_\text{H}}{a_0}  < 4.11\times 10^{-4}\text{Mpc}^{-1} (4.12 \times 10^{-4}),\;\; {\rm  at \;95\% \; C.L.}\nb\\
\eqn
When the BICEP/KECK 2014 data is included, the Planck TT + lowP +BK 14 (Planck TT, TE, EE + lowP + BK 14) data yields
\bqn
\frac{k_\text{H}}{a_0}  < 4.08\times 10^{-4}\text{Mpc}^{-1} (4.07 \times 10^{-4}),\;\; {\rm  at \;95\% \; C.L.}\nb\\
\eqn
These upper bounds show that the observational constraints on the pre-inflationary dynamics effects are robust to different data sets (without/with polarization data or BK 14 data included) and if the tensor spectrum is included.

Using the relation
\bqn
\frac{k_\text{H}}{a_0} =\frac{\sqrt{\gamma_{\rm B}}}{3} \frac{a_{\rm B}}{a_0} m_{\rm Pl}=\frac{\sqrt{\gamma_{\rm B}}}{3} m_{\rm Pl} e^{-N_{\rm tot}},
\eqn
where $N_{\rm tot} \equiv \ln{(a_0/a_{\rm B})}$ denotes the total e-folds from the quantum bounce until today, the above upper bounds on $k_{\rm H}/a_0$ can be translated into the constraint on the  total $e$-folds
$N_{\text{tot}} $ as
\bqn
N_{\rm tot} \gtrsim 140  \;\; (95\% {\rm C.L.}),
\eqn
where we have taken $\rho_\text{c}=0.41 m_\text{Pl}^4$. This in turn leads to a lower bound $\delta N_* > N_{\text{tot}}-N_*-N_\text{after}$, where
$\delta N_* \equiv \ln{(a_*/a_B)}$, $N_* \equiv \ln{(a_{\text{end}}/a_*)}$, and $N_\text{after} \equiv \ln{(a_0/a_{\text{end}})}$, where  $a_*$ denotes
 the expansion factor at the moment that our current horizon exited the Hubble horizon  during the slow-roll inflation, and $a_{\text{end}}$ is that at the end of inflation. Taking $N_*\simeq 60\simeq N_\text{after}$,
we find
\bq
\delta N_* \gtrsim 20.
\eq
With this constraint, one may think that if there are some modes at large scale could exit the Hubble horizon when the slow-roll inflation is not fully satisfied. However, according to the analysis in \cite{zhu_pre-inflationary_2017}, the number of e-folds before the onset of the slow-roll inflation is in general about $\mathcal{O}(5)$, therefore the number of e-folds of the slow-roll inflation has to exceed 60. This fact implies that all the relevant modes at observable scales exit the Hubble horizon when the slow-roll approximation is fulfilled.

\begin{table*}
\caption{Best fit values of the six cosmological parameters and the  constraints on $k_\text{H}/a_0$ and $r$ at 95\% C.L for different cosmological models from different data combinations.}
\lb{bestfit}
\begin{ruledtabular}
\begin{tabular} {ccccccc}

 Parameters  & Planck TT & Planck TT,TE,EE   & Planck TT & Planck TT,TE,EE  & Planck TT     &Planck TT,TE,EE\\
 \,\,        & +lowP     &+lowP              &+lowP+$r$  &+lowP+$r$         &+lowP+BK14+$r$ &+lowP+BK14+$r$\\
\hline
{\boldmath$\Omega_b h^2$} & $0.02242$ &0.02220 &0.02223 &0.02208 &0.02232 &0.02232\\

{\boldmath$\Omega_c h^2$} & $0.1181$ &0.1200 & 0.1208 &0.1209 &0.1205 &0.1193\\

{\boldmath$100\theta_{MC} $} & $1.04079$&1.04056 &1.04085 &1.04072 &1.04109 &1.04113 \\

{\boldmath$\tau      $} & $0.084$ &0.079 &0.079 &0.069 &0.081 &0.094\\

{\boldmath${\rm{ln}}(10^{10} A_s)$} & $3.101 $&$3.092$ &3.095 &3.074 &3.095 &3.118\\

{\boldmath$n_s            $} & $0.965   $&$0.964$ &0.963 &0.963 &0.967 &0.968\\
\hline
\boldmath{$k_{\rm H}/a_0$} & $<4.17\times 10^{-4}$ &$ <4.15\times 10^{-4}$ &$<4.11\times 10^{-4}$ &$< 4.12 \times 10^{-4}$ &$< 4.08 \times 10^{-4}$ &$< 4.07 \times 10^{-4}$\\
$ r$ &  $ ---- $& $---$ & $ < 0.110 $& $ < 0.106 $ &$ < 0.0650 $ &$ < 0.0655 $\\
\end{tabular}
\end{ruledtabular}
\end{table*}

\subsection{Implications on non-Gaussianity and power asymmetry}

Similar to the case in the dressed metric approach, the bouncing phase prior to the standard slow-roll inflation leads to excited states at the onset of the slow-roll inflation for certain perturbation modes with $k \lesssim k_{\rm H}$. These effects are encoded in the Bogoliubov coefficients $\alpha_k$ and $\beta_k$ (as given in (\ref{akbk})), and essentially depend on the parameter $k_{\rm H}$, and thus represent a characteristic feature of the hybrid quantization approach to cosmological perturbations in LQC. In general, as pointed out in \cite{agullo_non-gaussianities_2011}, the excited states that are different from adiabatic states at the onset of the slow-roll inflation could generate an enhancement on the non-Gaussianity of primordial perturbations in the squeezed configurations which involve very different scales. Recently, the impact of the excited states produced by quantum bounce in the dressed metric approach in LQC has been studied in \cite{agullo_loop_2015,zhu_primoridal_2018, agullo_non-gaussianity_2018}, and there are two main consequences. First, if the three different scales involved in the non-Gaussisnity are all in the observable range, the corrections on the non-Gaussisnity due to the quantum bounce effects are at the same magnitude of the slow-roll parameters and thus are well within current observational constraints. Second, if one of the scale is at superhorizon scale, then the non-Gaussianity in the squeezed limit can be enhanced and it is these effects that can yield a large statistical anisotropy on the power spectrum. As we mentioned in the above, the quantum gravitational  effects in the hybrid quantization approach are very similar to those in the dressed metric approach, thus it is natural to ask whether the above two consequences still hold. In this subsection, we are going to investigate  these points.   

To proceed, it is convenient to start with the amplitude $f_{\rm NL}$ of the non-Gaussisnity in the squeezed limit. As shown in \cite{zhu_primoridal_2018}, the contribution from the excited states takes the form,
\bqn
f_{\rm NL} \equiv \frac{\frac{5}{6} B_{\mathcal{R}} (k_1, k_2, k_3)}{P_{\mathcal{R}}(k_1) P_{\mathcal{R}}(k_2)+ P_{\mathcal{R}}(k_1)P_{\mathcal{R}}(k_3) + P_{\mathcal{R}}(k_2)P_{\mathcal{R}}(k_3)},\nb\\
\eqn
where $B_{\mathcal{R}}(k_1, k_2, k_3)$ denotes the bispectrum which characterizes the non-Gaussianity of the comoving curvature perturbations $\mathcal{R}$, $k_1, \; k_2, \; k_3$ are three scales involved in the non-Gaussianity, and
\bqn
P_{\mathcal{R}}(k) \equiv \left|\alpha_k+\beta_k\right|^2 \frac{H^2}{8 \pi^2 M_{\rm Pl}^2 \epsilon_1} \frac{2\pi^2}{k^3}.
\eqn
We are interest in the squeezed configuration which involves two very different scales, i.e., $k_2 \simeq k_3 \gg k_1$. Considering that $k_1$ is small enough compared to the other two modes, one finds,
\bqn
f_{\rm NL}^{\rm squeezed} &\simeq& \frac{10}{3}\epsilon_1 \left[\frac{k_3}{k_1}+\frac{9 k_1}{4 k_3} + \mathcal{O}\left(\frac{k_1^2}{k_3^3}\right)\right]\nb\\
&& \times {\rm Re}\Bigg[\frac{(\alpha_{k_1} +\beta_{k_1})(\alpha_{k_3}+\beta_{k_3})}{|\alpha_{k_1} +\beta_{k_1}|^2 |\alpha_{k_3}+\beta_{k_3}|^2}\nb\\
&&~~~~~~~~  \times (\alpha_{k_1}^* -\beta_{k_1}^* ) \alpha_{k_3}^* \beta_{k_3}^*\Bigg].
\eqn
In the observational range, $\frac{k_3}{k_1}$ could be as large as $\sim 10^4$. Thus in general one expects in the squeezed limit that the non-Gaussianity gets enhanced due to the excited state, i.e., $\beta_k \neq 0$. However, whether the non-Gaussianity gets enhanced or not also depends on the magnitude of the Bogoliubov coefficient $\beta_{k_3}$ for the smallest scale. In order to see the quantum bounce effects in this limit, let us study it in detail.

If all three scales ($k_2 \simeq k_3 \gg k_1$) involved are all in the observational range, in order to make $f_{\rm NL}^{\rm squeezed}$ larger,  a natural choice for $k_2 \simeq k_3 \gg k_1$ is $k_2 \simeq k_3 \gg k_{\rm H}$ and $k_1 \simeq k_{\rm H}$. If this is the case, the magnitude of $f_{\rm NL}^{\rm squeezed}$ will be mainly determined by $\beta_{k_3}^*$, which is exponentially suppressed as
\bqn
\beta_{k_3}^* \simeq i 2 \cos\left(\pi\sqrt{ \frac{5}{12}}\right) e^{- \pi \sqrt{ \frac{5}{6}} \frac{k}{k_{\rm H}}} + \mathcal{O}\left(e^{-2 \pi \sqrt{ \frac{5}{6}}\frac{k}{k_{\rm H}}}\right).\nb\\
\eqn
Then we infer from this expression that
\bqn
f_{\rm NL}^{\rm squeezed} < \epsilon_1 \times  \mathcal{O}(1).
\eqn
Similar to the case in the dressed metric approach,  the quantum bounce effects in the squeezed limit are strongly suppressed by the factor $e^{- \pi \sqrt{5/6} k/k_{\rm H}}$ even though $k_3/k_1 \gg 1$. Therefore, quantum gravitational effects on the non-Gaussianity only contribute to the same order as that in a slow-roll inflation of a  single field in the observable range, which are well within current observational constraints. We note that this property is expected and also agrees with the results obtained in the dressed metric approach both analytically \cite{zhu_primoridal_2018} and numerically \cite{agullo_non-gaussianity_2018}.

On the other hand, if one of the three scales is at superhorizon scale (i.e. $k_1 \ll k_{\rm H}$) and other two  are at observable scales ($k_2 \simeq k\simeq k_3$), then $f_{\rm NL}^{\rm squeezed}$ reads
\bqn
f_{\rm NL}^{\rm squeezed}(k) \simeq \mathcal{O}(1) \times \epsilon_1 \frac{k k_{\rm H}}{k_1^2}
 {\rm csch}\left(  \frac{\pi \sqrt{5} k }{\sqrt{6} k_{\rm H}}\right),
 \eqn
 which is suppressed by $ {\rm csch}\left(  \frac{\pi \sqrt{5} k }{\sqrt{6} k_{\rm H}}\right)$ at small scales ($k \gg k_{\rm H}$) but enhanced dramatically by the factor $k k_{\rm H}/k_1^2$ at large scales ($k \simeq k_{\rm H}$). This behavior agrees qualitatively with results in the dressed metric approach , but is quantificationally different \cite{agullo_non-gaussianity_2018, zhu_primoridal_2018}. Such enhancement can produce modulation on the primordial curvature power spectrum. In particular, due to the EKC mechanism \cite{erickcek_superhorizon_2008, erickcek_hemispherical_2008}, the superhorizon modes could bring modifications at observational scales, which is expected as an approximately linear function of positions. This could naturally provide an explanation to the observed power asymmetry in the CMB spectrum. According to the analysis given in \cite{lyth_cmb_2013, namjoo_hemispherical_2013}, the relation between the power asymmetry and non-Gaussianity is given by
 \bqn
 A(k)=\frac{6}{5} |f_{\rm NL}^{\rm squeezed}| k_1 x_{\rm cmb} \mathcal{P}^{1/2}_{\mathcal{R}}(k_1).
 \eqn
This equation is also known as the consistency condition, relating the amplitude of power asymmetry to the amplitude of the non-Gaussianity in the squeezed limit. Using the observational constraint on $k_{\rm H}/a_0$ obtained in the above section and since $A(k) \sim 0.066$ at large scales $(k/a_0)^{-1}  \sim 3 {\rm Gpc}$, one expects
\bqn
\frac{k_1}{a_0} \lesssim 10^{-6} {\rm Mpc}^{-1},
\eqn
in which we have used $x_{\rm cmb} = 14 {\rm Gpc}$. At small scales, similar to that in the dressed metric approach, the non-Gaussianity amplitude $f_{\rm NL}$ reduces to the usual magnitude with $f_{\rm NL} \sim n_s -1$. Thus, at small scales the power asymmetry is small, which is consistent with the constraint from quasars \cite{hirata_constraints_2009}.

\section{Conclusions}

In this paper,  we have provided a detailed and analytical study of the evolutions of the primordial perturbations during pre-inflationary phase and their observable effects on inflationary perturbation spectra for a single field inflation in LQC within the framework of the hybrid approach. Comparing to the dressed metric approach that the time-dependent mass of perturbation modes near the quantum bounce is negative for initial kinetic energy dominated conditions of inflaton field, the main discrepancy of the hybrid approach is that the effective time-dependent mass is positive for the same initial conditions. We show that it is this positive mass that leads to non-adiabatic evolution of perturbation modes near the  bounce at large scales,  which in turn generates excited states on the primordial comoving curvature perturbations rather than the usual BD vacuum state at the onset of the slow-roll inflation,  and leaves oscillating features on primordial perturbation spectra at large scales.  These features can be constrained by observational data and using the Planck 2015 temperature (TT+lowP), polarization (TT,TE,EE+lowP), and BICEP/KECK 2014 data, we found the upper bound for $k_{\rm H}/a_0 \lesssim 4.1  \times 10^{-4} {\rm Mpc}^{-1}$ at 95\% C.L., which provides a lower bound for the total $e$-folds from the quantum bounce until now as $N_{\rm tot} \gtrsim 140 \; (95 \% {\rm C.L.})$. With this constraint, we considered the impact of the quantum gravitational effects on the non-Gaussianities of the primordial curvature perturbations and  showed that the amplitude of the non-Gaussianity in the squeezed limit due to these effects is small at the observable scales, but can be enhanced if we consider a superhorizon mode that couples to the observable modes at large scales. It is this enhanced non-Gaussianity that leads to a modulation on the isotropic primordial power spectrum at the observed scales,  and thus can naturally provide an explanation of the power asymmetry observed in the CMB spectrum.

\section*{Acknowledgements}
The authors are grateful to Guillermo A. Mena Marug\'an for reading the manuscript and for useful comments. This work is supported in part by National Natural Science Foundation of China with the Grants Nos. 11675143 (Q.W. \& T.Z.)
 and 11675145 (A.W.).

\end{document}